\documentclass[twocolumn,floatfix,preprintnumbers,amsmath,amssymb]{revtex4}
\usepackage{epsfig}
\usepackage{color}
\usepackage{amsmath}
\usepackage{amssymb}
\usepackage{citesort}
\usepackage{psfrag}
\usepackage{pstricks}
\usepackage{longtable}
\usepackage{dcolumn}
\usepackage{bm}

\newcommand{\bra}[1]{\mathinner{\langle{#1}|}}
\newcommand{\ket}[1]{\mathinner{|{#1}\rangle}}

\newcommand{\braket}[2]{\langle #1|#2\rangle}

\providecommand{\abs}[1]{\lvert#1\rvert}

\begin{document}

\title{Engineering directed excitonic energy transfer}

\author{Alejandro Perdomo}
\affiliation{Department of Chemistry and Chemical Biology, Harvard University, 12 Oxford Street, 02138, Cambridge, MA}

\author{Leslie Vogt}
\affiliation{Department of Chemistry and Chemical Biology, Harvard University, 12 Oxford Street, 02138, Cambridge, MA}

\author{Ali Najmaie}
\affiliation{Department  of Chemistry and Chemical Biology, Harvard University, 12 Oxford Street, 02138, Cambridge, MA}

\author{Alan Aspuru-Guzik}
\affiliation{Department  of Chemistry and Chemical Biology, Harvard University, 12 Oxford Street, 02138, Cambridge, MA}
\email{aspuru@chemistry.harvard.edu}

\begin{abstract}
We provide an intuitive platform for engineering exciton transfer dynamics. We show that careful consideration of the spectral density, which describes the system-bath interaction, leads to opportunities to engineer the transfer of an exciton. Since excitons in nanostructures are proposed for use in quantum information processing and artificial photosynthetic designs, our approach paves the way for engineering a wide range of desired exciton dynamics. We carefully describe the validity of the model and use experimentally relevant material parameters to show counter-intuitive examples of a directed exciton transfer in a linear chain of quantum dots.
\end{abstract}


\maketitle


The widely-applied F\"orster theory for energy transfer links experimental results to estimates of system information, particularly in biological and nanoscale applications \cite{scholes2003,lakowicz_principles_2006}. The usefulness of this theory is partly due to the simple expression of the kinetic rate constants as a product of electronic coupling and a spectral overlap factor which captures the complexity of the environment. F\"orster theory describes transport in the incoherent limit, but a complementary and more elaborate approach, such as Redfield theory, is often required to describe energy transfer. However, the information essential to understanding the dynamics is buried within the structure of the equations.  In this letter, we employ a quantum kinetic rate approach to distill the information contained in equations into a simple, yet instructive, formula.  We use this approach to design directed exciton transfer mediated by an environment.

Excitonic energy transfer (EET) has been studied in systems as varied as quantum dot (QD) nanostructures \cite{scholes_excitons_2006,crooker2002}, polymer chains \cite{collini_coherent_2009}, and photosynthetic complexes \cite{lee_coherence_2007,cheng_dynamics_2009}.  Many applications of EET would benefit from controlling exciton dynamics. Perfect state transfer, as studied in the quantum computing community, is achievable in certain engineered systems, but only at particular times during coherent evolution \cite{PhysRevA.71.032312}.  Recent works have shown that environment-induced decoherence can alter exciton dynamics \cite{rozbicki_quantum_2008,rebentrost_environment-assisted_2009,plenio_dephasing-assisted_2008}, although controlling the transfer direction has only been achieved using external potentials \cite{high_control_2008}.  Our paper builds upon the idea of engineering exciton transfer by designing appropriate system-bath interactions \cite{gaab_effects_2004,Cao_Silbey_2009}.  We show that it is possible to design experimentally realizable systems where the environment can be used to direct the flow of energy.


The Hamiltonian used in our simulation aims to capture dynamics in a single-exciton manifold \cite{may_charge_2004} interacting with an environment,
\begin{equation}\label{eq:hamiltonian}
\hat{H} = \hat{H}_s + \hat{H}_b + \hat{H}_{sb}
\end{equation}
with
\begin{equation}\label{eq:Hs}
\hat{H}_s = \sum_{n=1}^N E_n \ket{s_n}\bra{s_n} + \sum_{n \neq m} J_{mn} \ket{s_m}\bra{s_n}.
\end{equation}
This representation is in the \textit{site basis} $\{\ket{s_n}\}$ of  localized excitations on each of $N$ sites, (e.g. QDs or chromophores), with excitation energy $E_n$ for each site and inter-site coupling $J_{mn}$. The environment is described by a phonon bath,
\begin{equation}\label{eq:Hb}
\hat{H}_{b} = \sum_{\mathbf{q}} \hbar \omega_{\mathbf{q}} (b^{\dagger}_{\mathbf{q}} b_{\mathbf{q}} + 1/2),
\end{equation}
where $b^{\dagger}_{\mathbf{q}}$ ($b_{\mathbf{q}}$) is the creation (destruction) operator for a phonon with wavevector $\mathbf{q}$. The system-bath interaction is assumed to be linear,
\begin{equation}\label{eq:Hsb}
\hat{H}_{sb} = \sum_{n=1}^N \ket{s_n}\bra{s_n} \sum_{\mathbf{q}} \hbar \omega_{\mathbf{q}} \Big(g^n_{\mathbf{q}} b^{\dagger}_{\mathbf{q}} + (g^{n}_{\mathbf{q}})^* b_{\mathbf{q}} \Big).
\end{equation}
where $g^n_{\mathbf{q}}$ describes the site-specific coupling of electronic and vibrational degrees of freedom.  We ignore the off-diagonal terms in the above equation, which correspond to phonon-induced modulations of the inter-site coupling, $J_{mn}$ \cite{may_charge_2004}. While fluctuations of the gap give rise to the diagonal electron-phonon coupling considered here  \cite{leegwater_coherent_1996}, the inter-site couplings are usually one or two orders of magnitude smaller than the excitation gap and are therefore kept constant. In general, the validity of this approximation is still an open question and its applicability varies from system to system.

To describe the excitonic quantum dynamics we use Redfield theory \cite{redfield1957,pollard_solution_1994,nitzan_chemical_2006}, which is a reduced density matrix approach in the regime of weak system-bath coupling. The formalism involves second-order perturbation theory in the system-bath interaction, $\hat{H}_{sb}$.  This method assumes the Markov approximation, no initial correlations between system and bath degrees of freedom, and a thermalized bath. We avoid the frequently employed secular approximation (Bloch equations) \cite{may_charge_2004,nitzan_chemical_2006}; it is important to note that for systems where the time scale $\abs{\omega_{ab} - \omega_{cd}}^{-1}$ is comparable or larger than the characteristic decoherence time, the coherence to population transfers contribute considerably to the dynamics of the system and must therefore be included.

The equation of motion for the density operator $\hat{\rho}(t)$ in the excitonic energy basis representation, $\hat{H}_s \ket{e_a} = \epsilon_a \ket{e_a}$, is given by \cite{redfield1957,pollard_solution_1994,nitzan_chemical_2006},
\begin{equation}\label{eq:master-equation}
\frac{\mathrm{d} \rho_{ab}(t)}{\mathrm{d}t} = -i \omega_{ab} \rho_{ab}(t) + \sum_{cd} R_{ab,cd} \rho_{cd}(t),
\end{equation}
with $\rho_{ab}(t) \equiv \bra{e_a} \hat{\rho}(t) \ket{e_b}$ and $\omega_{ab} = (\epsilon_a - \epsilon_b)/\hbar$.
The first term on the right hand side of Eq.~\ref{eq:master-equation} describes the fully coherent dynamics in the absence of $\hat{H}_{sb}$ and the second term describes the irreversible dynamics from the interaction with the phonon bath.

Correlations in bath-fluctuations on different sites \cite{richter2006} are taken into account by using the relation $g^n_{\mathbf{q}}g^{m*}_{\mathbf{q}} = g^m_{\mathbf{q}}g^{n*}_{\mathbf{q}} = g^2_{\mathbf{q}} e^{-R_{mn}/R_{\rm{corr}}}$, where $R_{mn} \equiv \abs{\mathbf{R}_{mn}}$ is the distance between the sites and $R_{\rm{corr}}$ is the phonon correlation length \cite{renger2002}.

Using this relation for the electron-phonon couplings, the cross-correlation $C_{mn}(\omega)$ can now be written as,
\begin{equation}\label{eq:Cw}
C_{mn}(\omega) = e^{-R_{mn}/R_{\rm{corr}}} C(\omega),
\end{equation}
where the frequency correlation function $C(\omega) = 2 \pi [n(\omega) + 1] (J(\omega)- J(-\omega))$, with $n(\omega)$ the Bose-Einstein distribution and the spectral density of the bath $J (\omega) = \sum_{\mathbf{q}} \abs{g_{\mathbf{q}}}^2 \omega_{\mathbf{q}}^2 \delta(\omega - \omega_{\mathbf{q}})$, where $\delta(\omega)$ is the Dirac delta function.

For the Hamiltonian specified in Eq.~\ref{eq:hamiltonian}, the Redfield tensor elements are given by
\begin{equation}\label{eq:Rtensor}
\begin{split}
R_{ab,cd} &= \Gamma_{db,ac}(\omega_{ca})+\Gamma^*_{ca,bd}(\omega_{db})\\
 & \quad -\delta_{bd} \sum_e \Gamma_{ae,ec}(\omega_{ce})- \delta_{ac} \sum_e \Gamma^*_{be,ed}(\omega_{de}),
\end{split}
\end{equation}
where $\delta_{ij}$ is the Kronecker delta and
\begin{equation}\label{eq:Gamma}
\begin{split}
\Gamma_{ab,cd}(\omega_{dc}) &= \frac{1}{2} \zeta_{ab,cd} C(\omega_{dc}) \\ & +
\frac{i}{2 \pi} \zeta_{ab,cd} \mathcal{P} \left( \int_{-\infty}^{\infty} \frac{C(\omega)}{\omega_{dc}-\omega} \, \textrm{d}\omega \right),
\end{split}
\end{equation}
\begin{equation}\label{eq:zeta}
\zeta_{ab,cd} = \sum_{n,m} (U^{-1})_{an} U_{nb} (U^{-1})_{cm} U_{md} e^{-R_{mn}/R_{\rm{corr}}},
\end{equation}
with $\mathcal{P}$ denoting the Cauchy principal value of the integral and $U_{na} = \braket{s_n}{e_a}$ the transformation matrix elements relating the site basis $\{\ket{s_n}\}$ and the excitonic basis $\{\ket{e_a}\}$. Since $\ket{e_a} = \sum_n U_{na} \ket{s_n}$, then $\abs{U_{na}}^2$ can be interpreted as the contribution of the $n$-th site to the $a$-th eigenstate of $\hat{H}_s$.


Hereafter, we will focus on QDs at low temperature (10 K) as our prototypical experimental realization. We neglect possible inversion asymmetry of the crystal, and therefore the contribution of the piezoelectric coupling, and focus instead on the deformation potential coupling. As shown by Calarco \textit{et al.} \cite{calarco2003}, the spectral density describing this coupling (in the absence of an external electric field) is given by
\begin{equation}
J (\omega) = \Theta(\omega) \eta \omega^3 e^{-\omega^2/\omega^2_c},
\end{equation}
where, $\Theta(\omega)$ is the Heaviside step function. We use typical values for GaAs QDs for our numerical simulations \cite{experimental_values,madelung_otfried_semiconductors:_2004}, giving $\eta = \frac{(D_e - D_h)^2}{4 \pi^2 \rho u^5 \hbar} = 0.035$  ps$^2$  and $ \omega_c = \sqrt{2 u^2/l^2} = 1.41$ ps$^{-1}$, where $D_e$ ($D_h$) is the deformation coupling potential for electron (hole), $u$ the speed of sound within the quantum dot, $\rho$ its mass density and $l$ the ground state localization length, assumed to be the same for electron and holes. We also assume a correlation length, $R_{\rm{corr}}$, of 3 nm.

Using Eqns.~\ref{eq:Rtensor}-\ref{eq:zeta}, the population transfer rates, $k_{ab}$, between the eigenstates $a \rightarrow b$ is given by,
\begin{equation}\label{eq:rate}
k_{ab} = R_{bb,aa} = \zeta_{ab,ba} C(\omega_{ab}).
\end{equation}
This equation is central to our insight into designing excitonic transfer. Though derived for a different regime, the form of Eq.~\ref{eq:rate} is similar to the widely used rate equation in F\"orster theory (incoherent limit), $k^{Forster}_{ab} \propto J^2_{ab} I_{ab}$, where $J$ is the electronic coupling, and $I$ is the spectral overlap integral \cite{scholes_long-range_2003}. Both equations are a product of two terms: one dealing primarily with the description of the system ($\zeta$ in this paper or $J^2$ for F\"orster theory), and another largely depending on and arising from the system-bath interaction ($C(\omega)$ in this paper or $I$ for F\"orster theory). F\"orster theory has been applied in fluorescence resonance energy transfer (FRET) to design chromophores for biosensing assays \cite{lakowicz_principles_2006,rogach_energy_2009,medintz_quantum_2009} and was recently verified experimentally for semiconductor QDs \cite{kim_experimental_2008}. We use the simple structure of Eq.~\ref{eq:rate} to gain microscopic and experimentally relevant insight into engineering directed and optimized EET.

While $C(\omega_{ab})$ depends on the overlap of system eigenenergies with the spectral properties of the phonon bath (e.g., lattice vibrations, solvent, protein environment, etc.), $\zeta$ depends on the transformation matrix $U$ in Eq.~\ref{eq:zeta}, determined by the relative magnitude of electronic couplings with respect to site energies, site connectivity, and spatial correlation between sites. The aim is then to maximize (minimize) the product of these two factors in Eq.~\ref{eq:rate} to favor (suppress) the desired rates.

To illustrate the applicability of Eq.~\ref{eq:rate}, we choose two three-site examples to highlight the importance of the phonon bath interaction to achieve directed EET.  Since multiplying a system Hamiltonian by a scalar does not change the maximum exciton transfer probability in the coherent limit, the distances and site energies for the two cases are chosen so that the Hamiltonians are related by a multiplicative factor ($3.5 \hat{H}^{\mathbf{1}}_{s} = \hat{H}^{\mathbf{2}}_{s}$).  As a consequence, the $\zeta$ values are roughly the same for cases \textbf{1} and \textbf{2} (Table~\ref{table:factors}); slight differences are introduced by the bath correlation term in Eq.~\ref{eq:zeta}.  In the cases chosen, the fully coherent evolution gives a maximum probability of finding an excitation of 5\% (1\%) for site 1(2).  Therefore, any difference between the two examples is a result of interaction with the environment.

In contrast to the $\zeta$ factors, the $C(\omega_{ab})$ values for cases \textbf{1} and \textbf{2} at 10 K differ by at least one order of magnitude due the position of the transition frequencies with respect to the spectral density (Table~\ref{table:factors}).  Any changes to the system Hamiltonian can affect $C(\omega_{ab})$.  In our examples, the scalar factor mentioned above changes the $C(\omega_{ab})$ such that the largest population transfer rate is switched from site 1 to site 2 since $\omega_{31}^{\bf{1}} \approx \omega_{32}^{\bf{2}}$ and $C(\omega^{\bf{1}}_{31}) \approx C(\omega^{\bf{2}}_{32})$ (Fig.~\ref{fig:examples}).

\begin{table}[h]
 \caption{\label{table:factors} Contributions of the system factor, $\zeta_{ab,ba}$, and of the overlap between transition frequency and phonon bath spectral properties, $C(\omega_{ab})$, to the calculation of the quantum kinetic rates $k_{ab}$ from energy eigenstate $\ket{e_a}$ to  $\ket{e_b}$. The two cases considered are described in Fig.~\ref{fig:examples}.}
\begin{tabular}{c @{\;\;\;\;\;\;} ccc @{\;\;\;\;\;\;} ccc}
\hline \hline
                                                            & \multicolumn{3}{c}{\bfseries Case 1} & \multicolumn{3}{c}{\bfseries Case 2} \\
  \hline
  $\ket{e_a} \rightarrow \ket{e_b}$ & $\log \zeta$ & $\log C$ & $\log k$ & $\log \zeta$ & $\log C$ & $\log k$ \\
  \hline
  $3 \rightarrow 1$                            & -1.5               & 11.6         & \bf{10.0} & -1.5              & 7.9           & \bf{6.3} \\

   $3 \rightarrow 2$                           & -2.4               & 10.7         & \bf{8.3}    & -2.4             & 11.6         & \bf{9.1} \\

   $2 \rightarrow 1$                           & -4.3              & 11.4          & \bf{7.1}    & -4.3             & 10.2         & \bf{5.8} \\
\hline \hline
\end{tabular}
\end{table}

From the values of $\zeta$ and $C(\omega_{ab})$, it is clear that case \textbf{1} is designed such that an excitation starting on site 3 will tend to transfer to site 1, but in case \textbf{2} the population will go to site 2, albeit at different rates. Simulations of the quantum dynamics according to Redfield theory confirms this result (Fig.~\ref{fig:results}). Over the typical exciton lifetime of 1 ns in QDs, we not only achieve directed transfer, but also population enhancement compared to both the maximum site population during fully coherent dynamics and the population expected at thermal equilibrium.

Moreover, while there are always experimental limitations in tuning parameters, the structure of Eq.~\ref{eq:rate} is valuable since it partitions the effects due to the system and bath. Using the calculated rates and/or a visual inspection of $C(\omega)$ (Fig. 1B), it is easy to determine the impact of varying a system parameter on exciton transfer.  Future work will address exciton and electron transfer between sites with varying spectral density functions, as well as the role of aligning dipole moment orientations in engineering EET. We are also working to identify regimes in which preserved coherences enhance or reduce the efficiency of excitonic transfer. Of course, in situations where multiple excitons are present in the system due to incident light intensity, frequency range, and/or optical spectral density of the quantum dots, the Hamiltonian used to describe the system needs to be expanded accordingly \cite{klimov_spectral_2007}. To the extent that each exciton couples to the environment through a spectral density as described in this letter, some of the intuition developed here should be transferable to these systems. However, there are a number of interesting complications, including many-body interactions among the excitons and corrections to the rates due to multi-phonon processes. A more careful analysis is needed when the offset in QD excitation energies is large compared with the characteristic frequency expanded by the phonon spectral densities or for systems at much higher temperature, e.g., room temperature, where multi-phonon processes are expected to be relevant, or even dominant. These effects are currently under investigation and are beyond the scope of the current communication.


In summary, we develop a framework for engineering environment-assisted and directed excitonic transfer in a network of coupled QDs based on a quantum kinetic rate approach. We emphasize the importance of how characteristic frequencies of a system fit within the spectral bath structure. Our examples utilize the factored and intuitive form of the population transfer rates equation, which separates the contributions from the system (electronic) and bath (vibrational) degrees of freedom. This equation is similar in spirit to the rate equation for FRET, making it convenient to design interesting scenarios for environment-assisted transfer. Although we focus on QD examples, the principles presented here form the basis for engineering a wide range of desired EET in a variety of nanostructures or artificial molecular photosynthetic units.

The authors thank Semion Saikin for helpful discussions.  A. A.-G. and A. P. were supported as part of the Center for Excitonics, an Energy Frontier Research Center funded by the U.S. Department of Energy, Office of Science, Office of Basic Energy Sciences under Award Number DE-SC0001088. L. V. acknowledges support from the NSF Graduate Research Fellowship. A.N. acknowledges support from NSERC (Canada)


\begin{thebibliography}{30}
\bibitem[{exp()}]{experimental_values}
\bibinfo{note}{We consider a typical GaAs QD with deformation potentials $D_e =
  -14.6$ eV, $D_h = -4.8$ eV, mass density of $\rho = 5.4$ g/cm$^3$, speed of
  sound $u = 5000$ cm/s, and radius of $l =5$ nm}.
\end{thebibliography}

\begin{newpage}
\begin{turnpage}
\begin{figure}[h]
\begin{center}
 \includegraphics[width=12.0 cm]{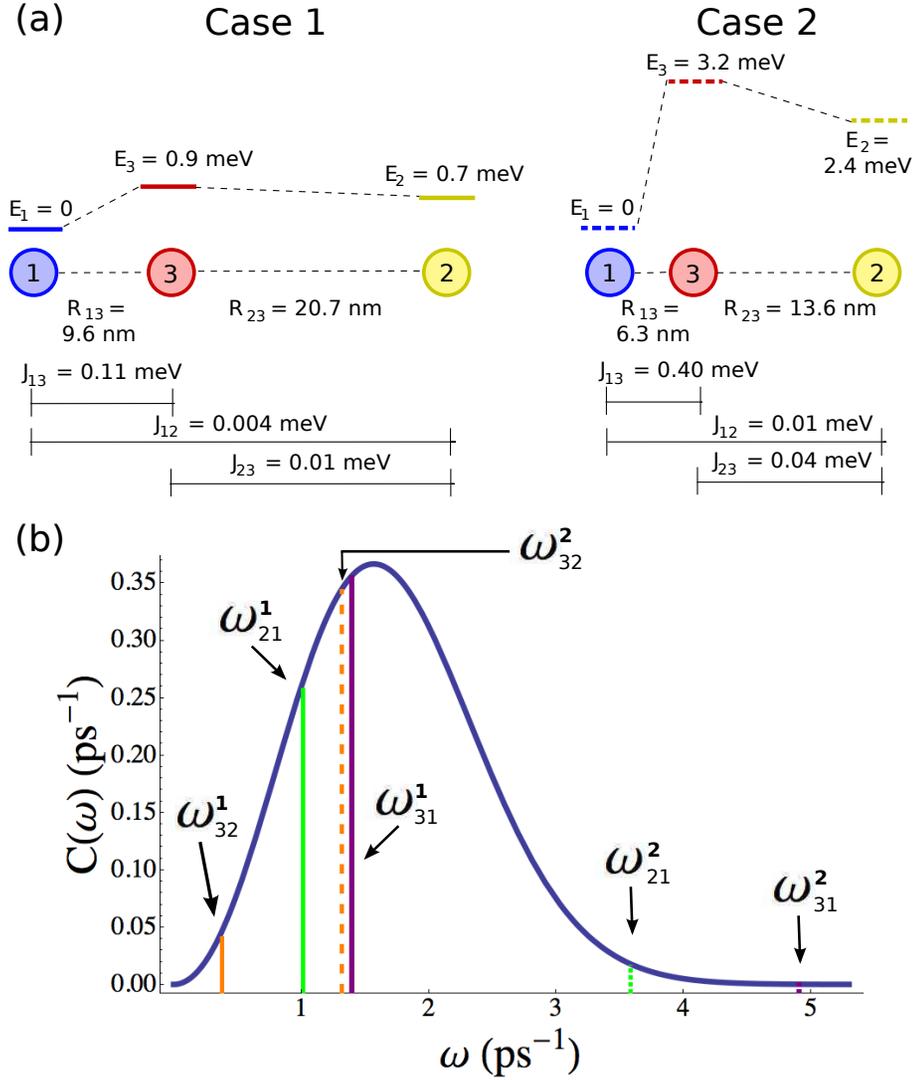}
\end{center}
\caption{\label{fig:examples}
Two cases of a three-site system are considered. a) Scaled schematic of system spacing and energy levels with details of variable site energies ($E_n$) and intersite F\"orster coupling strengths ($J_{mn}$). For QDs with transition dipole moments aligned perpendicular to $R_{mn}$, $J_{mn} =100 \,\textrm{meV}/R^3_{mn}$, with $R_{mn}$ in nanometers \cite{nazir_anticrossings_2005}. b) Frequency correlation function for a super-ohmic spectral density.  Energy basis transition frequencies for case \textbf{1} (\textbf{2}) are indicated by solid (dashed) vertical lines.
}
\end{figure}
\end{turnpage}
\end{newpage}

\begin{turnpage}
\begin{figure}[h]
\begin{center}
 \includegraphics[width=19.0 cm]{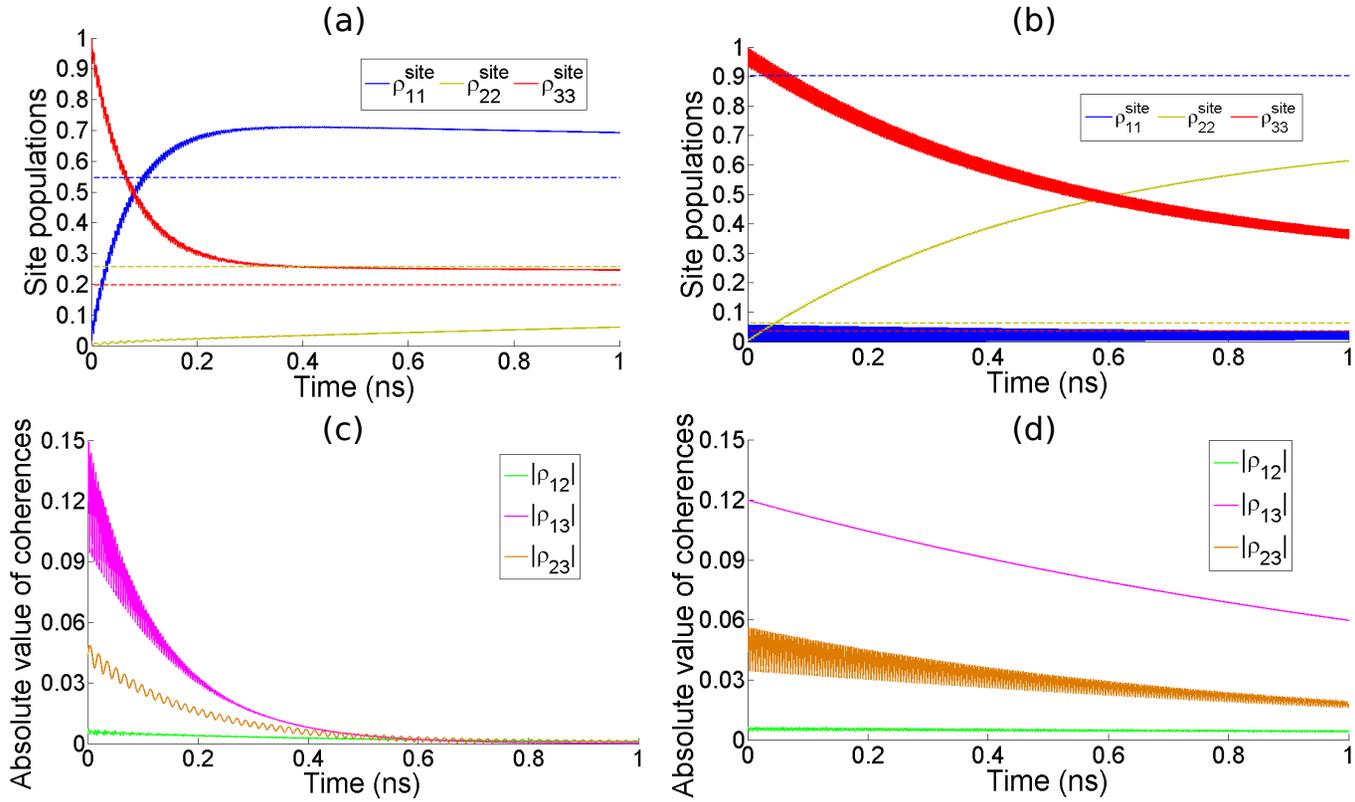}
\end{center}
\caption{\label{fig:results}
Site basis population probabilities for an excition starting on site 3 for case \textbf{1} (a) and case \textbf{2} (b) demonstrate the change in transfer dynamics obtained by scaling the Hamiltonian (simulation at 10 K).  Dashed lines indicate site populations at thermal equilibrium.  Energy basis coherences for case \textbf{1} (c) and case \textbf{2} (d) display characteristic oscillations and damping over the course of 1 ns, a typical recombination time in QD systems.
}
\end{figure}
\end{turnpage}

\end{document}